\documentclass[conference,compsoc]{IEEEtran}
\usepackage{amsmath,amssymb,amsfonts,amsthm}
\usepackage{algorithmic}
\usepackage{textcomp}
\usepackage{xcolor}
\usepackage{colortbl}
\usepackage{multirow}
\usepackage{url}
\usepackage{mdframed}

\newtheorem*{definition}{Definition}
\surroundwithmdframed{definition}
\theoremstyle{definition}

\ifCLASSOPTIONcompsoc
  \usepackage[nocompress]{cite}
\else
  \usepackage{cite}
\fi
\ifCLASSINFOpdf
  \usepackage[pdftex]{graphicx}
\else
\fi
\usepackage{amsmath}

\usepackage{tabularx}

\ifCLASSOPTIONcompsoc
  \usepackage[caption=false,font=footnotesize,labelfont=sf,textfont=sf]{subfig}
\else
  \usepackage[caption=false,font=footnotesize]{subfig}
\fi
\begin{document}
%
\title{Towards the Definition of Enterprise Architecture Debts}

\author{
\IEEEauthorblockN{Simon Hacks\IEEEauthorrefmark{1},
Hendrik H\"ofert\IEEEauthorrefmark{2},
Johannes Salentin\IEEEauthorrefmark{2},
Yoon Chow Yeong\IEEEauthorrefmark{3}, and
Horst Lichter\IEEEauthorrefmark{1}
}
\IEEEauthorblockA{\IEEEauthorrefmark{1}Research Group Software Construction\\
RWTH Aachen University,
Aachen, Germany\\
\textit{\{hacks,lichter\}@swc.rwth-aachen.de}}
\IEEEauthorblockA{\IEEEauthorrefmark{2}
RWTH Aachen University,
Aachen, Germany\\
\textit{\{hendrik.hoefert,johannes.salentin\}@rwth-aachen.de}}
\IEEEauthorblockA{\IEEEauthorrefmark{3}
Universiti Teknologi Petronas\\Perak Darul Ridzuan, Malaysia\\
\textit{yeongyoonchow@gmail.com}}
}


%


\maketitle

\begin{abstract}
In the software development industry, technical debt is regarded as a critical issue in term of the negative consequences such as increased software development cost, low product quality, decreased maintainability, and slowed progress to the long-term success of developing software. However, despite the vast research contributions in technical debt management for software engineering, the idea of technical debt fails to provide a holistic consideration to include both IT and business aspects.

Further, implementing an enterprise architecture (EA) project might not always be a success due to uncertainty and unavailability of resources. Therefore, we relate the consequences of EA implementation failure with a new metaphor --Enterprise Architecture Debt (EA Debt). We anticipate that the accumulation of EA Debt will negatively influence EA quality, also expose the business into risk. 
\end{abstract}

\begin{IEEEkeywords}
Enterprise Architecture Management, Enterprise Architecture Debt (EA Debt), Term Definition
\end{IEEEkeywords}

\IEEEpeerreviewmaketitle

\section{Introduction}
\label{sec1:introduction}
Technical Debt is a metaphor that had been introduced by Cunningham \cite{Cunningham1993}. In the software development industry, technical debt is regarded as a critical issue in term of the negative consequences such as increased software development cost, low product quality, decreased maintainability, and slowed progress to the long-term success of developing software \cite{Tom2013}. Technical debt describes the delayed technical development activities for getting short-term payoffs such as a timely release of a specific software \cite{Zazworka2011PrioritizingOpportunities}. Seaman et al. \cite{Seaman2012} described technical debt as a situation in which software developers accept compromises in one dimension to meet an urgent demand in another dimension and eventually resulted in higher costs to restore the health of the system in future.

Furthermore, technical debt is explained as the effect of immature software artifacts, which requires extra effort on software maintenance in the future \cite{Guo2011AManagement}. The concept of technical debt reflects technical compromises that provide short-term benefit by sacrificing the long-term health of a software system \cite{Li2015AManagement}. In view of the original idea of technical debt that focused on the code level in software implementation, the concept had been extended to software architecture, documentation, requirements, and testing \cite{Brown2010}. While the technical debt metaphor has further extended to include database design debt, which describes the immature database design decisions \cite{Albarak2018}, the context of technical debt is still limited to the technological aspects. 


There is extensive literature which have been done on the concept of technical debt which can be sub-categorized into design debt, (software) architectural debt, code debt, documentation debt, etc. However, the concept of technical debt that particularly focuses on technical aspects demonstrates a lack of attention to attaining a holistic perspective to address the alignment between business and IT aspects. While enterprise architecture management (EAM) is gaining significant attention as a management instrument in business and IT \cite{Lange2016}, to the best of our knowledge, there is no existing research on the metaphor of ``Enterprise Architecture Debt''. While technical debt focuses primarily on the effect of immature software artifacts, which leads to a huge amount of cost in the future \cite{Guo2016CostsStudy,Guo2011AManagement,Ampatzoglou2015}, the concept of EA Debt extends its focus to include business aspects. Since business and IT representatives potentially have different mindset and different goals \cite{Kuznetcov2014}, it is believed that EA Debt plays an important role to provide a common language for them.


For the purpose of introducing the metaphor, we formulate following our research questions. We consider our main goal in \emph{RQ1}, with its two major sub-questions \emph{RQ1.1} and \emph{RQ1.2} which have to be answered in order to answer the primary research question:
\\\\
(\emph{RQ1}) \emph{What is an adequate definition of Enterprise Architecture Debt?} 
\\\\
(\emph{RQ1.1}) \emph{How is Technical Debt defined and what criteria is related to it?}
    \\\\
(\emph{RQ1.2}) \emph{Which aspects are important for the quality of EAs and which layers are included in it?}
    \\

In this paper, we argue for an enterprise architecture perspective to expand the concept of technical debt to encompass business and IT aspects for a more holistic view. The core contribution of this study is a new proposed metaphor --Enterprise Architecture Debt-- that addresses the limitation of technical debt in providing a comprehensive enterprise architectural view. Mapping the debt concept to enterprise architecture is also a remarkable initiative to deal with EA implementation tactfully.    

The remainder of this paper is organized into the following parts: Section~\ref{sec2:methodology} describes the research method; Section~\ref{sec:concepts} introduces the concepts of technical debt and enterprise architecture; those concepts are facilitated to come to a definition of enterprise architecture debt in Section~\ref{sec:definition}, while we present two examples for EA Debt in Section~\ref{sec:examples};
lastly, we express implications and future works in Section~\ref{sec6:futurework}, and concluded with Section~\ref{sec7:conclusion}.

\section{Research Method}
\label{sec2:methodology}
This work employs a Design Science Research Methodology (DSRM) which is developed by \cite{Peffers2007}. The core idea of design science research is to create and evaluate an innovative and purposeful artifact that enable organizations to address an important and relevant problem \cite{Hevner2004}. In order to systematically address the defined research questions, this study follows the six core activities defined in DSRM by \cite{Peffers2007}, as described following. 

\textbf{Activity 1: Problem Identification and Motivation.}
A problem discovery activity is mandatory to identify the research domain and potential subject to be addressed. 
The extant literature reveals a weakness of the technical debt concept in addressing business aspects and, thus, this drives our initiative to extend the metaphor of technical debt and make it applicable in enterprise architecture context. When existing studies suggested that technical debt brings negative impacts if it is left unmanaged \cite{Zazworka2011PrioritizingOpportunities, Brown2010}, we recognize that the accumulation of EA Debt brings danger to enterprise as well.

\textbf{Activity 2: Define the Objectives of a Solution.}
The research objectives and goals are formalized to solve the identified problem and, consequently, provide the intended direction for the research. This work aims to introduce the metaphor of enterprise architecture debt as well as to provide theoretical understanding regarding EA Debt. In addition to the theoretical contribution, this work also intends to conceptually understand EA Debt items from the perspective of four enterprise architectural levels which are described by TOGAF \cite{TOGAF}. The four architectural layers --business architecture, technology architecture, application architecture and data architecture introduced by TOGAF \cite{TOGAF}-- help to ensure EA Debt items are recognized from a comprehensive perspective, instead of focusing primarily on technical aspects. 

\textbf{Activity 3: Design and Development.}
The artifacts developed as the outcome of this activity demonstrate the contribution of this research to the academic and EA practitioner communities in which the concept and taxonomy of EA Debt metaphor will be developed. To develop the EA Debt metaphor, we elaborate on the relevant properties of EA, take a closer look at the definition of technical debt, and, then, join the findings.

\textbf{Activity 4: Demonstration.}
After creating the artifact, its use is demonstrated by solving one or several instances of the problem. We conduct two fictitious case studies to show how our definition can support enterprise architects by identifying improvement potentials within the EA as well as providing argumentation aids to steer the development of the EA towards its desired state.

\textbf{Activity 5: Evaluation.}
In the evaluation, the researcher observes and measures how well the artifact supports the solution of the problem. However, the evaluation of our definition is beyond the scope of this work. For future work, further detailed evaluation activities are expected to be carried out in the form of real case studies and expert interviews may be conducted with EA practitioners to validate the usage of the proposed approach.

\textbf{Activity 6: Communication.}
Towards the end of DSRM, it is mandatory to document the problems, solution, objectives, description related to the developed artifact for communication with the relevant audience. The outcome of research is presented in the form of research paper and insights are given to provide a direction for future research. 


\section{Key Concepts and Related Work}
\label{sec:concepts}

Before we deduce the term ``EA Debt'', we elaborate more on the terms Technical Debt (Section~\ref{subsec:technical debt}) as well as Enterprise Architecture and its quality issues (Section~\ref{subsec:enterprise architecture}).

\subsection{Technical Debt}
\label{subsec:technical debt}
The metaphor \emph{Technical Debt} was first introduced by Cunningham \cite{Cunningham1993} and mentions what we today would call ``refactoring''. 
This first idea of not-quite-right code which we postpone making it right, is expanded by various people to display also other kinds of debts or ills of software development, such as test debt, people debt, architectural debt, requirement debt, documentation debt, or an encompassing software debt \cite{Kruchten1, Seaman2011}.

According to Kruchten et al. \cite{Kruchten1} Technical Debt refers to invisible elements, because visible elements for improving, like new features for evolution or repairing defects for maintainability issues, should not be considered as debt. 
Technical debt should rather serve as a retrospect reflecting change of the environment, rapid success, or technological advancements as a possible cause for debt. However, ``the debt might actually be a good investment, but it's imperative to remain aware of this debt and the increased friction it will impose on the development team'' \cite{Kruchten1}. Hence, tools are required to increase the awareness to identify debt and its causes, and to manage debt-related tasks. Finally, the debt should not be treated in isolation from the visible elements of evolving and maintaining. Consequently, debt is ``the invisible result of past decisions about software that negatively affect its future'' \cite{Kruchten1}.

Tufano et al. \cite{Tufano} encountered the same phenomenon and point out that most code smells are introduced at their creation. Furthermore, the code often gets smellier due to new artifacts being build on top of suboptimal implementations. Even refactoring is often done wrong as it introduces further bad smells, which highlights the need for techniques and tools to support such processes \cite{Tufano}.

McConnell \cite{McConnell}, for example, tried to categorize different types of Technical Debt. He stresses that with this metaphor business and technical viewpoints can be emphasized, so that communication regarding specific problems can be enhanced. Technical debt is used as a uniform communication tool, that allows us to measure and keep track of debt which eventually should help find a suitable solution to the upcoming challenges. In this case it should also reflect the different viewpoints, including the stakeholder's perspective, in order to allow an effective collaboration \cite{McConnell, Theodoropoulos}.

Fowler even came up with a categorization of Technical Debt with his ``Technical Debt Quadrant'' to identify different types of it \cite{Fowler2}. He distinguishes technical debt into reckless / prudent and deliberate / inadvertent debt. This reflects the different scenarios where debt is taken and hints at debt being taken unconsciously or negligent sometimes.


For him, the metaphor's primary task is to help ``thinking about how to deal with design problems, and how to communicate that thinking'' \cite{Fowler2}. Nevertheless, it is used as a tool, which can reveal possible drawbacks of a current design decisions: his differentiation makes everyone aware that a certain decision could cause debt. This theoretical concept should then help to find a reasonable solution \cite{Power}.

Further, Technical Debt tries to help to decide how to invest scarce resources: ``Like financial debt, sometimes technical debt can be necessary'' \cite{Brown}. Most of the time this debt is not visible, as \cite{Kruchten1} also pointed out, leading to making debt visible as one purpose. Additionally, the value and present value play a role, including the difference between the actual state and an supposed ideal state as well as the time-to-impact. This involves a differentiation between ``structural issues (the potential technical debt) and the effect it has on actual development (the effective technical debt)'' \cite{Schmid}, which could also be called problems and risks.

Overall debt has to be seen in its environment and it has to be determined if the debt was strategic or unintentional. Ultimately, this Technical Debt can be considered as an external software attribute, which needs to be quantified \cite{Brown,Lavazza}. As a consequence, one has to measure all criteria to estimate the debt and make a proper decision based on that information. It has been shown that reasonable decisions can be made more easily, if corresponding information (debt) is taken into account. Additionally, delaying a supposed ``right'' implementation has a significant impact on the cost of the project, so that an appropriate management of the debt concept is useful \cite{Codabux, Guo, Oliveira}.

\subsection{Enterprise Architecture}
\label{subsec:enterprise architecture}
Following, we focus on \emph{Enterprise Architecture} and its quality issues. In accordance to ISO/IEC 25010 quality ``is the degree to which a product or system can be used by specific users to meet their needs to achieve specific goals with effectiveness, efficiency, freedom from risk and satisfaction in specific contexts of use'' \cite{ISO25010}.

As described by Saint-Louis et al. \cite{SaintLouis} the definitions for EA diverge significantly, already hinting at differences in quality issues. Kappelman et al. \cite{Kappelman} pointed out that ``The `enterprise' portion of EA is understood by some as a synonym to `enterprise systems', yet by others as equivalent to `business' or `organization'. [...] Even less uniform is the understanding of the meaning of `architecture'. The most common understanding of the term is a collection of artifacts (models, descriptions, etc.) that define the standards of how the enterprise should function or provide an as-is model of the enterprise'' \cite{Kappelman}. However, EA helps to face ``ongoing and disruptive change'' \cite{SaintLouis} by attempting to align IT and business strategy. 

Despite the diverse perspectives and definitions, some articles focus on quality issues regarding EA. One could for example derive qualities that encompass IT system qualities, business qualities, and IT governance qualities \cite{Saat}. 
Henderson and Venkatraman's Strategic Alignment Model (SAM) identifies Business Strategy, Business Structure, IT Strategy and IT Structure as four key domains of strategic choice, the so called ``alignment domains'' \cite{Henderson}. Based on this information an artifact-based framework for business-IT misalignment symptom detection was created in the article of the same name \cite{Dora}. This framework also includes a first suggestion of catalogues for misalignment symptoms, EA artifacts, and EA analysis methods, so that also rather invisible or underestimated elements are considered. Moreover, a link between those categories is established, in order to know which misalignment symptoms affect which artifacts and how this can be analyzed.

Addicks and Appelrath \cite{Addicks} searched for key figures and their metrics in order to unitize the quality assessment of an application landscape. This approach can be applied to the EA as a whole, because business processes for example influence the application's quality. They stated that ``all key figures must at least fit one of the following three conditions: (a) it must be used for indications of applications and be based on the application’s attributes, (b) it must be an indicator of an application and its value is determined by attributes and relations from other EA artifacts (the applications’ enterprise context), and (c) it must indicate a landscape’s quality and therefore use all attributes of applications and their enterprise context'' \cite{Addicks}.

Since many aspects influence the system's properties a unified meta model is difficult to create. However, Saat et al. \cite{Saat} show that even different enterprise orientations with divergent focuses prefer certain qualities over others. In general, they show that the striven qualities differ across enterprises, nevertheless, there are some general qualities that are desired in most situations. Thus, a specialized prioritization and adaptation is needed for the EA and its IT/business alignment.



More general approaches are followed by for example artifact based viewpoints of Winter and Fischer \cite{Winter} or the TOGAF Standard \cite{TOGAF}. They refer to domains like Business / Process Architecture, Software / Data Architecture, Application / Integration Architecture and Technology / Infrastructure Architecture.

Ylimäki goes even further and defines twelve critical success factors for Enterprise Architecture. These factors obviously influence EA and its quality, although they are different to previously known aspects. Here high-quality EA is described with: ``high-quality EA conforms to the agreed and fully understood business requirements, fits for its purpose (e.g. a more efficient IT decision making), and satisfies the key stakeholder groups’ (the top management, IT management, architects, IT developers, and so forth) expectations in a cost-effective way understanding both their current needs and future requirements'' \cite{Ylimaki}.

Moreover, there is an Enterprise Architecture Model Quality Framework (EAQF)
proposed by Timm et al. \cite{Timm}, which builds upon six principles to assess the quality of EA models, namely the principles of validity, relevance, clarity, economic efficiency, systematic model construction and comparison. Each of the principles is augmented with quality attributes that can be assessed in order to determine the quality addressing the EA Model's purpose, a specific view of it and the overall EA Model \cite{Timm}.

The EA purpose, its objectives and the stakeholders' concerns affect the EA model's quality assessment. The EA model as a whole and each EA model view on its own are important to the quality, such that the interaction of multiple model views focusing on different issues should not be underestimated. Although this framework targets the quality of EA models and not the EA directly, it provides reasonable attributes for its quality assessment.

\section{Defining Enterprise Architecture Debt}
\label{sec:definition}
As seen in Section~\ref{subsec:technical debt}, the definitions of Technical Debt are mostly descriptive, naming properties and different types of debt, rather than explicitly defining the term. Nevertheless, Technical Debt is considered as a tool pointing at possible risks, measuring and tracking deficits, and aiming at supporting the process of finding suitable solutions. The focus mainly lies on increasing awareness of also invisible or structural elements and giving a uniform basis for discussions and communication.

To outline the findings regarding Enterprise Architecture and its quality issues from Section~\ref{subsec:enterprise architecture}, we conclude that there are different definitions of Enterprise Architecture around. Although it is more often described and its quality aspects are mentioned with respect to a specific viewpoint and enterprise orientation, the descriptions diverge. Hence, only a common understanding of EA and a basis for communication and discussion is established.

Therefore, the requirements differ for each enterprise, making a uniform approach according to EA models and quality issues for a certain level of detail impossible. Consequently, our definition of Enterprise Architecture Debt determines a technique providing some crucial factors in order to estimate an EA's quality for its specific purpose on a high abstraction level and increase the awareness for possible suboptimal aspects that may cause severe drawbacks in the future. The metaphor of debt should serve as a common basis for communication and discussion, as well as pointing out differences in as-is situations and proposed ideal to-be situations \cite{Kappelman}. The need for such a basis was observed before as \cite{Niu} identified four possible situations regarding communication:

\begin{enumerate}
    \item Consensus where stakeholders use terminology and concepts in the same way.
    \item Correspondence where they use different terminology for the same concepts.
    \item Conflict where they use the same terminology for different concepts.
    \item Contrast where stakeholders differ in terminology and concepts.
\end{enumerate}

Furthermore, flexibility and robustness are identified as important factors in the dynamically changing and evolving environment of the enterprise and its structures. Hence, it can be used as a powerful tool, like Technical Debt, and help to manage a complex enterprise, because it allows to get a holistic overview and keep track of existing debt. All in all this should fulfill the purpose and function of EA according to Kappelman et al. \cite{Kappelman}.

In order to achieve a scenario-unaware definition of Enterprise Architecture Debt, we have to state what we refer to when we are talking about EA. \cite{SaintLouis} conducted a SLR (Systematic Literature Review) and found multiple definitions of EA. Therefore, we regard EA as a set of artifacts, which are aggregated.

In the TOGAF Standard version 9.2 Business Architecture, Data Architecture, Application Architecture and Technology Architecture are identified as the four architecture domains \cite{TOGAF}. Those roughly match the different layers of \cite{Winter}, although they named Process Architecture in particular.

The artifacts themselves and their importance for a specific enterprise may differ, so we focus on the following aspects in particular, which are general enough to be applied to the majority of EAs:


\begin{center}
    \renewcommand{\labelenumii}{\theenumii}
    \renewcommand{\theenumii}{\theenumi.\arabic{enumii}.}
    \begin{enumerate}
        \item Enterprise Architecture layers
        \begin{enumerate}
            \item Business Architecture
            \item Process Architecture
            \item Software Architecture and Used Services (SaaS)
            \item Technology Architecture (Hardware)
        \end{enumerate}
        \item Other influencing factors
        \begin{enumerate}
            \item Stakeholders
            \item Guidelines and Standards
        \end{enumerate}
    \end{enumerate}
\end{center}

Considering EA as a set of artifacts, it can contain more than the mentioned artifacts, but it is easy to add and remove artifacts later, so the metaphor of Enterprise Architecture Debt can be adapted to individual situations. 

Winter and Fischer \cite{Winter} point out that ``Most of the artifacts [...] in EA can be represented as aggregation hierarchies''. Taking this aggregation hierarchies and the definition of Technical Debt by Cunningham \cite{Cunningham1993} into account, EA Debt in general can be understood as an aggregation of taking debts in each layer. But, just adding up each artifact would not give a concrete overview of the current situation, it could even whitewash huge issues in the whole EA. Therefore, every artifact, and also every part an artifact consists of, should be weighted. Obviously, there is no uniform weighting function, because it depends on the concrete EA and also the organization. On top of that, we need to take into consideration that the interrelations along artifacts can cause debt, too. An artifact might be optimized and work perfectly, but if there is a huge overhead caused by interfaces respectively multiple lines of reporting, the EA might not work perfectly.

According to Hurley \cite{Hurley} a definition consists of ``the definiendum'', the word to be defined, and ``the definiens'', the words that do the defining. Those again can be split into ``generic elements'' and ``specific elements'' or ``characteristics''. Accordingly, our definition reads as follows:

\begin{definition}[EA Debt]
    Enterprise Architecture Debt is a \textbf{\emph{metric that depicts the deviation}} of the \textbf{\emph{currently present state}} of an enterprise from a \textbf{\emph{hypothetical ideal state}}.
\end{definition}

Based on this definition, we can explain and characterize appropriate objectives and details:

Enterprise Architecture Debt arises, when debt is taken in an artifact, which an EA consists of. This means that an element is not implemented or executed optimally in relation to the supposed ideal situation. Taking debt in a low hierarchy, can be helpful and pay off, but it has to be ``repaid'' as fast as possible. Otherwise the whole EA would rely on that debt and use faulty or considered bad artifacts. This entails a high risk of additional debt and hinders the development. EA Debt is further increased by bad interfaces or bad interoperability and different priorities of stakeholders, not conform with an EA that is considered good by evaluation approaches.

Here a focus on mainly invisible elements can be helpful, because these factors may not even be recognized or their impacts are underestimated. By increasing the awareness for such issues the overall inadvertent debt can be reduced. Assuming a prudent management, this would lead to debt mostly being taken consciously and planned strategically. This means that possible repercussions are weighed and less trade-offs are accepted, which unnoticeably impair the enterprise.

Hence, the identified differences between the current state and the supposed ideal situation can be managed, such that the sometimes necessary debt, namely deliberate prudent debt, holds the largest share. 


\section{Demonstrating EA Debt}
\label{subsec:evaluating EA Debt}
Applied standards and guidelines are an indication for reliable artifacts and therefore lower EA Debt. The EAQF \cite{Timm} assesses the quality of an EA model with six principles. These principles can be used to construct an ideal situation and detect the differences to the current as-is situation. In this way also rather invisible or unconsidered elements can be found, which then form a good entry point for EA Debt estimations. Yet again, the awareness of invisible or hardly noticeable elements has to be increased, because they in particular entail high risk of further debt in the future.

\H{O}ri \cite{Dora} proposed an artifact-based framework that aims at detecting misalignment in an undesired state of the enterprise, which can be used to detect also EA Debt. Based on the strategic alignment perspectives Strategy Execution, Technology Transformation, Competitive Potential, and Service Level the framework decomposes those perspectives into corresponding perspective components, namely the ``alignment matches''. Then, they are connected to typical misalignment symptoms, using a misalignment symptom catalogue as a reference. After that, relevant containing artifacts are identified, again using an artifact catalogue. Finally, suitable EA analysis types are collected respecting the affected artifacts. This article already sets up catalogues based on other research that can be used as a reference \cite{Dora}.

Despite those frameworks present helpful approaches to evaluate EA and its models, there are also some shortcomings. They can be referred to EA itself or to our new idea of EA Debt. As mentioned already before, the development of a suitable ideal situation for an enterprise is still a hard task. Furthermore, there exists no uniform solution, so that the transfer from one organization to another can be quite challenging.

Some observations are made by Schmid \cite{Schmid} regarding shortcomings of Technical Debt that can be transformed and extended to EA Debt: ``Technical debt should be evaluated with respect to future evolution. [...] We need to differentiate between the structural issues (the potential technical debt) and the effect it has on actual development (the effective technical debt). [...] There is nothing like a technical-debt-free system.'' In general for the entire organization and our concept of Enterprise Architecture Debt this implies:

\begin{enumerate}
    \item EA Debt should be evaluated with respect to future evolution, because further development may rest upon the current sub-optimal implementations and structure.
    \item We need to differentiate between the structural issues (the potential EA Debt) and the effect it has on actual development (the effective EA Debt). One could also refer to problems and risks that can have diverging impacts on the EA. Even though something is not implemented in the best possible way it may not have a severe effect on the quality.
    \item There is nothing like an EA-debt-free system. So the hypothetical ideal state will never be reached. Only reducing debt can underline a good development, as long as the debt was valued correctly.
\end{enumerate}

\subsection{Possible impacts on EA Debt}
\label{subsec:impacts on EA Debt}

As mentioned before there are multiple artifacts that have an impact on EA Debt, identifying them should be an outcome of EA Debt evaluation. Having a look at the development of artifacts over time \cite{Addicks}, known to cause EA Debt can be one way to approach the search of new impacts to the current EA Debt situation. Finding a slow or even stagnating development in an artifact $A$ means there is almost no effort to lower the debt and, therefore, artifacts relying on $A$ will increase the overall EA Debt. Starting a search at $A$ can help finding impacts on EA Debt. On top of the already mentioned artifacts, that are able to cause EA Debt, there are more hidden or invisible aspects. We propose some of them: 
\begin{enumerate}
    \item Communication overhead (documentation)
    \item Interface bottlenecks (incompatibility)
    \item Contradictory goals of stakeholders
    \item Integrity problems or information inconsistency
\end{enumerate}
Adapted to individual EAs there can be many more, these are just some artifacts that can have additional impact on EA Debt.

Another impact, that has to be taken into consideration is that EA Debt itself should not be ignored and deferring the evaluation can, but not has to, increase EA Debt exponentially. Ideally, everyone working on artifacts of an EA should at least know the concept, so everyone is aware that his acting can cause EA Debt.

\subsection{Examples}
\label{sec:examples}
We come up with two made up and simplified examples, one showing where taking EA Debt is useful and one showing that EA Debt is lowering the efficiency of the whole enterprise and should be removed as fast as possible.

\subsubsection{Example 1 -- Necessary EA Debt}
\label{subsec:example1}
A company is situated in the insurance market and implemented its business critical applications on their mainframe until the end of the last decade. Due to a change in their IT-strategy, future applications should be developed using cloud environments. As the application landscape is comprised by more than 300 applications, a big bang scenario, where all applications are moved to the cloud, is unfeasible. Consequently, the applications are moved to the cloud step by step according to their application life-cycle rating.

The central enterprise architecture department has defined a target landscape and a road map describing the way to get to the target landscape. This road map includes also two applications $a$ and $b$, which should be moved from the mainframe to the cloud. Both applications depend on each other, which means that they use interfaces of each other. As it is planned that both applications should be moved to the cloud simultaneous, the interfaces of both applications can be developed within the target landscape. 

Due to unforeseeable delays in the project that implements application $b$, it is not expected that $a$ and $b$ can still go live at the same time in the cloud. However, the interfaces of $b$ are indispensable for the use of $a$. Therefore, $a$ is developed in a way that it relies on the interfaces of the mainframe implementation of $b$. As this interface implementation is obviously not part of the target landscape, this will introduce EA Debts into the EA.

Nonetheless, there is no feasible alternative and, therefore, there is the need to take this additional effort, which leads to a worse quality of the overall EA. However, the concept of EA Debt can help to create awareness for this quality issue along all stakeholders and that this EA Debt should be repaid as soon as application $b$ has been moved to the cloud.

\subsubsection{Example 2 - Unnecessary EA Debt}
\label{subsec:example2}
Next, we consider the same company as in the example before. The company employs several different servers to host their web-based applications. The security operations team is heavily occupied and, therefore, they prioritize all notifications for new security issues of their systems.

For example, there is one notification for a minor security issue of their web-servers. As this issue demands a significant effort to be fixed, the operations team decides not to fix this issue and to focus on other issues. This will create an EA Debt as this lowers the overall quality of the EA and, consequently, should be considered to be repaid as soon as possible.

Taking debt in this case is different from the first example in Section~\ref{subsec:example1}, because the enterprise has no other chance then to take the debts. Here in the second example (Section~\ref{subsec:example2}) the enterprise can decide to invest directly or to take the debt and maybe repay it later. Thus, EA Debt also emphasizes this risk.

\subsubsection{Findings}
In two made up and simplified examples we showed, that EA Debt can be a tool, like monetary debt, to grow, but it can also be dangerous, if it potentially can not be repaid and consequently does not pay off. In contrast to monetary debt, EA Debt can suddenly occur, but does not have to. This means that enterprises have to be careful with taking EA Debt and even more with not lowering it. Still EA Debt points at those risks and problems to facilitate a sustainable development. Another example, illustrating an as-is situation with EA Debt in an EA would be rather uninteresting, because there are no universal guidelines yet, how to deal with EA Debt (see future work, Section~\ref{sec6:futurework}) but every enterprise has to deal with EA Debt themselves. Nevertheless lowering EA Debt should be the main goal in such a situation. With the two examples, we demonstrate a way, how to use the concept of EA Debt as a tool to prevent bad decisions regarding the whole EA.

\section{Implications and Future Work}
\label{sec6:futurework}
This section discusses the constructive implications for real-life practice and future research, as well as the limitations found in this study. The objective of this session is to provide a future direction for real-world application and further research which has the potential to highly contribute to the enterprise architecture domain. 

By introducing a new metaphor that addresses a current gap in technical debt, the theoretical and empirical contribution can in turn benefit both the academia and EA practitioners, respectively, for research and real-life practices. The contribution of our work implies:
\begin{enumerate}
    \item A realization of enterprise architecture debt concept for EA practitioners, IT representatives, and management. It allows EA practitioners to conceptualize the components that might affect the success of EA implementation;
    \item A tool for EA practitioners to critically identify and examine enterprise architecture debts across four architectural layers on the basis of TOGAF;
    \item A method for enterprise architects to effectively communicate the EA problems to management for the success of EA implementation; 
    \item A new research area for practices, approaches, models, or tools relevant to the context of EA Debt, such as EA Debt measurement, EA Debt identification, EA Debt monitoring, etc.
\end{enumerate}

In practice, several existing methods can be used to facilitate the identification, detection and measurement of EA Debt to base the estimates. EA Debt can be identified and measured by evaluating the performance of EA models using existing EA analysis tools proposed by Buschle et al. \cite{Buschle2011}. Moreover, Enterprise Coherence Assessment (ECA) proposed by Wagter et al. \cite{Wagter} can be used as an instrument to measure enterprise's level of coherence during enterprise transformation and thus incoherence between important aspects of the enterprise can be identified as EA Debt. Also, \H{O}ri \cite{Ori2017} and Carvalho and Sousa \cite{Carvalho2008} proposed approaches for business-IT misalignment detection. Nevertheless, future work would suggest to develop a comprehensive framework, which is unique to the context of EA Debt.

This study is limited to describing the proposed approach in terms of the measurement units and process steps, as well as a preliminary validation based on synthetic case study, but not a full validation in real settings. The complete validation can be done through a real case study or expert interviews in future research. With the case study results within different organizations, even more discussion can be provided on its practical applicability in future. From there, further refinements of the proposed approach can be made in order to improve its quality and flexibility.

\section{Conclusion}
\label{sec7:conclusion}
Implementing enterprise architecture is essential to enhance the business-IT alignment in a holistic manner. However, academia, software developers, and organizations have been focusing on technical debt, which deals with the quality issues on code, application and system level. Considering the importance of EA in creating value to organizations, this work has explored a new metaphor to extend the benefits of technical debt concept, which is Enterprise Architecture Debt. In the pragmatic world, EA projects, programs or initiatives are implemented to realize the target architecture and EA Debt is expected to incur along the EA implementation process due to limited resources. We have asserted that the accumulation of EA Debt over time is likely to negatively affect the quality of an enterprise architecture in responding to the complex business environment. 

As we opened with the definition of EA Debt a new area of research, future work can elaborate on several different topics. First, the overall concept of EA Debt needs to be further evaluated, due to case studies, surveys, and so on. Second, it should be investigated how existing approaches can contribute to this new field, like EA quality assessments, EA best practices, etc. Third, existing concepts of code smell detection can be transferred to the domain of EA Debts to create an initial catalog of potential EA smells. Fourth, research should be conducted to identify EA smells, which solely arise in the field of EA and have no counterpart in existing Technical Debt domains. Last, research can elaborate on management methods for EA Debt, like how to decide which EA debt to repay next or how to involve EA stakeholders.

%




\bibliographystyle{IEEEtran}

%



\end{document}